\documentclass[openacc]{rsproca_new}
\usepackage{upgreek}
\usepackage{bm,amssymb,amsmath}
\usepackage{bbm}
\usepackage[polutonikogreek, english]{babel}
\usepackage[english]{betababel}

\usepackage{wasysym}
\usepackage[numbers]{natbib}
\usepackage{epsfig}
\usepackage{pstricks}
\usepackage{lpic}
\usepackage{etoolbox}

\newtheorem{theorem}{\bf Theorem}[section]

\newtheorem{corollary}{\bf Corollary}[section]
\newtheorem{definition}{Definition}[section]

\usepackage[normalem]{ulem}

\begin{document}

\title{A local diagnostic energy to study diabatic effects on a  class of degenerate Hamiltonian systems with application to mixing in stratified flows}

\author{
A. Scotti$^{1}$ and P.-Y. Passaggia$^{1}$}

\address{$^{1}$ Dept. of Marine Sciences, University of North Carolina at Chapel Hill, Chapel Hill, NC 27599}

\subject{Oceanography, Fluid dynamics}


\corres{A. Scotti\\
\email{ascotti@unc.edu}}

\begin{abstract}
In Hamiltonian systems characterized by a degenerate Poisson algebra, we show how to construct a local energy-like quantity that can be used to study diabatic effects on the evolution of the Available Energy of the system, the latter concept formalizing the original idea of Margules'.  
We calculate the local diagnostic energy for geophysically relevant flows. For the particular case of stratified Boussinesq flows, we show that under moderately general conditions, in inertial frames where the initial distribution of potential vorticity is even around the origin, our framework recovers the Available Potential Energy introduced by Holliday and McIntyre \cite{HollidayM81}, and as such depends only on the mass distribution of the flow. 
In non-inertial frames, we show that the local diagnostic energy of flows which are, in an appropriate sense, characterized by a low-Rossby number ${\rm Ro}$ ground state, has to lowest order in ${\rm Ro}$, a universal character. 

\end{abstract}


\begin{fmtext}
\section{Introduction}

A major  {challenge} in contemporary oceanography is to understand the role of small scale turbulence in regulating the oceanic Meridional Overturning Circulation (MOC), the process that over
millennial time scales exchanges surface with deep water \citep{Lozier15}. Given the large storage capacity of the deep ocean for heat and greenhouse gases, 
understanding the drivers of the MOC is essential for climate prediction \citep[see, e.g][and references therein]{Lynch17}. 
\citet{MunkW98} used energetics arguments {in an attempt} to quantify the amount of energy required to sustain the MOC,
but significant gaps remain in quantifying the pathways that energy injected
\end{fmtext} 
\maketitle
\noindent 
 at large scales 
by winds and tides take to reach the small 
scales at which mixing occurs \citep{WunschF04}.
Particularly vexing is the problem of estimating the energetic cost incurred when  turbulent processes irreversibly mix the stratifying agents 
\citep{IveyWK08}, that is how to estimate the energy input necessary to sustain a given rate of dissipation of the variance of the stratifying agents.


For a fluid in the Boussinesq approximation, \citet{WintersLRD95} used the concept of Available Potential Energy (APE) {\em as a diagnostic tool} to achieve such a relation. APE has a long history, starting with the pioneering work of \citet{Gibbs:73} and \citet{Margules03}, later developed by \citet{Lorenz55}  into a working tool, the so-called Lorenz Energy Cycle, which is still used today \citep{Storchetal12}. Essentially, Margules' idea was to calculate a minimum-energy state{,} compatible with certain constraints, and use the energy of such state to "gauge" the amount of potential energy effectively available {to produce mechanical work}, yielding a definition of the APE of the system.

Winters {\it et al.} considered the {global} effects of mixing (i.e., diabatic effects) on a closed system. For this purpose, they only needed a definition of APE for the system as a whole. There are however situations that call for a definition of APE that apply to localized regions of the domain, e.g. when measuring the  energy carried by nonlinear internal waves \citep{ScottiBB06,Lamb08}, when partitioning energy between  mean and {fluctuating} components in a cyclone \citep{KucharskiT00}, in determining  the efficiency of different mixing systems \citep{ScottiW14}, and in studying mixing and turbulence in spatially inhomogeneous systems \citep{PassaggiaSW17}. In all these studies, the starting point was  the local definitions of APE developed in the early 1980's by \citet{HollidayM81} for incompressible flows and by \citet{Andrews81} for compressible flows, based on a reference state that depends only on the mass distribution, though \citet{Andrews81} already suggested that more general reference states can be considered. Indeed, in the atmospheric literature, more general reference states have been considered \citep[e.g.,][]{CodobanS03,Andrews06}, whereas in the oceanographic applications the Lorenz paradigm still dominates \citep{Storchetal12,ZemskovaWS15}. 
For a recent review see \citep{Tailleux13}. 

The inadequacy of the standard definition of energy as  the sum of kinetic, potential and, for compressible fluids, internal energy to quantify the capacity of the system to do actual work, is connected to the degeneracy of the Poisson algebra in  the Hamiltonian formulation  of the problem \citep{Littlejohn82}. By degeneracy, we mean that the center of the Poisson algebra contains non-constant functionals of the phase space, the so-called Casimir functionals, or Casimirs for short. The set of Casimirs is an ideal of the algebra, and thus can be used to define a notion of equivalence
on the set of Hamiltonians:  Two Hamiltonians are equivalent, in the sense that they give the same dynamics, if they differ by a Casimir.  In other words, the Hamiltonian possesses a local (in phase space) gauge symmetry. 
From this point of view,   
the local APE formulations can be seen as selecting, out of a specific equivalence class, an Hamiltonian that satisfies one or more additional conditions \citep{Shepherd93}, given by a gauge-fixing condition.  As we shall clarify later, the Casimir includes the effects of constraints on the system. 

In this paper, we aim to revisit the issue of constructing a local diagnostic energy that can be used to diagnose the effect of diabatic processes on the energetics of fluid systems that, in the adiabatic limit, are described by a Hamiltonian with a degenerate Poisson algebra. {In particular}  {we} are seeking a quantity with the following properties:
\begin{enumerate}
\item Accounts for relevant  constraints.
\item Locality, and such that it satisfies (up to diabatic effects) suitable conservation laws.
\item Can be connected to Margules' intuitive notion of Available Energy when the latter is properly formalized.
\item Convexity in phase space, so that it can be meaningfully partitioned into a mean and eddy (or turbulent) component.
\item Its evolution under diabatic conditions reflects the loss of Available Energy. 
\end{enumerate}
Following \citet{Shepherd93}, we first lay down in sec.~\ref{sec:theory} the general theoretical framework that applies to generic systems that in the adiabatic limit have a Hamiltonian description characterized by a degenerate Poisson algebra. 
From there, we introduce the specific gauge-fixing condition that, when applied to the equivalence class of Hamiltonians, identifies the one whose density is the local diagnostic energy that we seek. 
In practice, the gauge-fixing condition identifies which Casimir needs to the added to the energy. At the same time, we also obtain the equations that specify the appropriate reference state. 

We then consider  two models which are commonly used in oceanography and which have an adiabatic limit described by a degenerate Hamiltonian structure: the incompressible shallow water equations and the incompressible Euler equations in the Boussinesq approximation for a continuously stratified flow. Both models are considered in inertial and non-inertial (i.e., rotating) frames.  For these systems, we give general properties for both the gauge-fixed Casimir and the reference state. In simple geometric configurations, we calculate analytically the solution or a suitable approximation. An interesting result that applies to both shallow water and Boussinesq equations is that, in rotating frames, the local diagnostic energy associated to low Rossby number reference states has a universal character. 

\section{Theoretical background}\label{sec:theory}
The phase space which describes a Hamiltonian mechanical system characterized by a finite number of degrees of freedom typically, but not always \cite[see, e.g.][p. 371--378]{Arnold13}, has a non-degenerate Poisson algebra. In this case, if no other constraints exist, two Hamiltonians are dynamically indistinguishable, and thus belong to the same equivalence class, if they differ by at most a time-dependent constant. We can think of this as a global symmetry. 

If, however, the Poisson algebra is degenerate, then the equivalence class, as pointed out in the introduction, is wider. If that is the case,  the global symmetry, manifested by the invariance of the dynamics under the addition of  a (time varying) constant to the energy at every point  in phase space, widens to a local (in phase space) symmetry, i.e. the energy can be altered by the addition of a Casimir that can depend on the location in phase space, and still yield the same dynamical equations. From a field-theoretical point of view, the Hamiltonian in this case possesses a local gauge symmetry \citep[see][for a simple, finite dimensional example, of a degenerate Hamiltonian system]{Littlejohn82}. 

It is important to remark that the existence of non-trivial Casimirs depends on the degeneracy of the Poisson algebra, {\em rather than} on the particular choice of the Hamiltonian. They represent conserved quantities whose existence does not depend via  Noether's theorem on symmetries of the Hamiltonian. If the algebra is degenerate, equilibrium solutions need not be (and in general, are not)  extrema of the Hamiltonian. 

 
For notational purposes, here and thereafter we will denote with $H[q]$ the "naive" Hamiltonian, i.e. the sum of kinetic, potential and, when needed, internal energy, while ${\cal H}[q]$ will denote a  member of the class of equivalent Hamiltonians, which can be written as ${\cal H}[q]=H[q]+C[q]$, the sum of the naive Hamiltonian and a Casimir. 
 It is also convenient to introduce the following notation: if $F[q]$ is a functional over the phase space which can be represented as the integral of an appropriate density $n-$form $\mathfrak{F}(q)$ over the $n-$dimensional oriented manifold $\mathbb{D}$, then $F(q)\equiv \star\mathfrak{F}(q)$ is the corresponding scalar density ($\star$ is the Hodge-star operator), i.e.
\begin{equation}
F[q]=\int_\mathbb{D}\mathfrak{F}(q)=\int_\mathbb{D} F(q)\mathfrak{V},
\end{equation}
where $\mathfrak{V}$ is the volume $n-$form. To summarize, square brackets $[\,]$ denote a volume integrated quantity, with the corresponding local density indicated by the use of $()$. 
In fluid systems \citep[for a review of the Hamiltonian formulation of fluid dynamics, see][and references therein]{Salmon88}, we find that 
 the Casimirs are generally underdetermined.  A set of Casimirs can be constructed as follows \citep[see][for an exhaustive compilation of Hamiltonians and associated Casimirs associated to geophysically relevant flows]{Shepherd90,AbarbanelHMR86}: Let $s_1(\mathbf{x}),\cdots,s_p(\mathbf{x})$ be $p$ Lagrangian invariants, by which we mean they are  intensive quantities ($0-$forms) that are constant along Lagrangian trajectories, i.e. 
\begin{equation}
\frac{\partial s_i}{\partial t}+{\cal L}_{\bf v}(s_i)=0,\,i=1,\ldots,p,\label{eq:C2.1}
\end{equation}
with ${\cal L}_{\mathbf v}$ being the Lie derivative w.r.t. the flow induced by $\mathbf{v}$. 
We also need a conserved density $\mathfrak{D}$, that is an $n$-form 
  representing a materially conserved extensive property, i.e.   such that 
\begin{equation}
\frac{\partial\mathfrak{D}}{\partial t}+{\cal L}_{\mathbf v}(\mathfrak{D})=0.
\label{eq:C2.2}\end{equation}
 For our purposes, we will use the mass $n-$form or, for incompressible flows, the volume $n-$form.   
Then 
\begin{equation}
C[q]=\int_{\mathbb{D}} C(s_1(\mathbf{x}),\cdots,s_p(\mathbf{x}))\mathfrak{D},\label{eq:C3}
\end{equation}
where $C(s_1,\cdots,s_p)$ is any real-valued function which depends on its arguments algebraically (i.e., we are not allowing $C$ to depend on derivatives of the $s_i$'s).
It is trivial to verify that $\mathrm{d}C/\mathrm{d}t\cong 0$ (here and thereafter $\cong$ means equality up to boundary terms, or, in the case of $n-$forms, equality up to an exact form). Note that there may be more general classes of Casimirs (e.g., which may depend on derivatives of the conserved quantities). Here we limit to Casimirs that can be written as (\ref{eq:C3}).

\subsection{Holonomic brakes and Available Energy}
It is straightforward to reformulate Margules' intuitive idea of "Available Energy" within the framework of a Hamiltonian system with a degenerate Poisson algebra. 

Consider a system described at time $t=0$ by a point $q_0$ in phase space. Under adiabatic conditions, the system is described by a trajectory $q(t)$ in phase space such that along the trajectory ${\cal H}[q(t)]={\cal H}[q_0]$, where ${\cal H}[q]$ is any member of the class of equivalent Hamiltonian functionals. 

While \citet{Gay-BalmazH13} modified the Hamiltonian dynamics to dissipate the Casimirs, here we introduce into the system what we may call a holonomic brake, whose purpose is to extract energy from the system (hence a brake) in such a way that the $s_i$'s
remain Lagrangian invariants (holonomic), and $\mathfrak{D}$ remains an integral invariant, i.e. (\ref{eq:C2.1}) and (\ref{eq:C2.2}) are still satisfied. Thus, the holonomic brake leaves the Casimirs {invariant}. Later, we will consider how holonomic brakes can be realized for particular classes of flows.


With the brake turned on, $\mathrm{d}{\cal H}[q(t)]/\mathrm{d}t\leq 0$   and, since the brake is holonomic, any decrease in ${\cal H}[q(t)]$ is due, entirely,   to a decrease in the "naive" Hamiltonian $H[q(t)]$. We assume that, in the limit $t\to\infty$, the application of the brake causes the system to relax to a stationary "ground" state $q_*$. If such a ground state exists and is unique,  the energy extracted by the holonomic brake from the system which is initially at $q_0$, is the difference $E_{\rm AE}[q_0]\equiv H[q_0]-H[q_*]$. We identify $E_{\rm AE}$ with the Available Energy envisioned by Margules. We have thus the following definition:
\begin{definition}
The Available Energy $E_{\rm AE}$ is the energy of the system referenced to the energy of the ground state, the latter being the state that the system settles to when subject to a holonomic brake.  
\end{definition}
We shall see presently that, in the process of determining $q_*$, a gauge-fixing condition will naturally arise which selects an element ${\cal H}_*$ of  the equivalence class of Hamiltonians. The rate of change in time of ${\cal H}_*[q(t)]$ measures the loss of Available Energy when $q(t)$ evolves under full diabatic conditions. 

\subsection{A special family of Casimirs: the Mass Distribution Function}
The main postulate of our approach is that the $q_*$ to which the system relaxes after the holonomic brake is turned on, is the point that minimizes the naive Hamiltonian subject to a set of constraints required by  the conservation of the $s_i$'s along trajectories, which are not affected by the brake.  Thus, the extremal point of $H[q]$ must be sought among the points that satisfy an extra set of conditions, the definition of which is the subject of this section. 

\begin{figure}
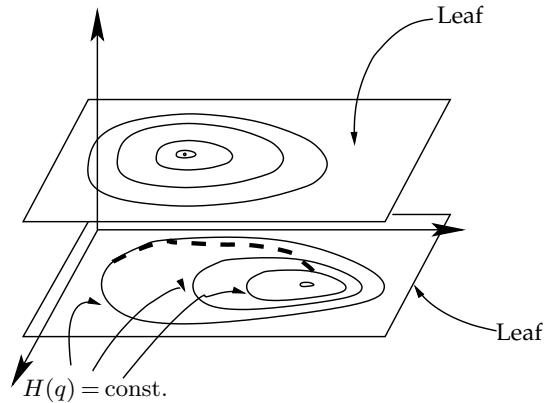

\begin{center}
\begin{lpic}{Phase_space(0.6)}
\lbl{99,83;Leaf}
\lbl{112,13;Leaf}
\lbl{19,0; $H(q)=\mathrm{const.}$}
\end{lpic}
\end{center}
\caption{\label{fig:1} Sketch of phase space with two leaves. The dashed line shows a trajectory under the action of a holonomic brake.  }
\end{figure}

Consider a Lagrangian set of coordinates ${\bm \alpha}=(\alpha^1,\ldots,\alpha^n)$.
A Lagrangian particle characterized by a $p$-tuple ${\bm s}=(s_1,\ldots,s_p)$ of Lagrangian Invariants retain its identity. The mass of all Lagrangian particles whose $p$-tuple falls between ${\bm s}$ and ${\bm s}+\mathrm{d}{\bm s}$ is given by
\begin{equation}
\Delta M({\bm s})=\left(\int_{\mathbb{D}}\left[\prod_{i=1}^p\updelta(s_i({\bm \alpha})-s_i)\right]\mathfrak{D}(\bm \alpha)\right)\mathrm{d}{\bm s},\label{eq:C_diff} 
\end{equation}
where $\mathfrak{D}(\bm \alpha)$ is the mass  $n$-form expressed in Lagrangian coordinates, and $\updelta$ is the Dirac's delta (not to be confused with the $\delta$ used to denote variations). Integrating $\Delta M$ over the $p-$tuples, from $s_i$ to its maximum value $S_i\equiv \max\{s_i(\bm \alpha), \bm \alpha\in \mathbb{D}\}$ for each $i$, we obtain the  Mass Distribution Function  (MDF)
\begin{equation}
M({\bm s})=\int_{s_1}^{S_1}\cdots\int_{s_p}^{S_p}\Delta M({\bm s}')=\int_{\mathbb{D}}\left[\prod_{i=1}^p\theta(s_i({\bm \alpha})-s_i)\right]\mathfrak{D}(\bm \alpha),
\label{eq:C_int}\end{equation}
where  
\begin{equation}
\theta(s)=\int_{\infty}^s\updelta(t)\mathrm{d}t,
\end{equation}
is the Heaviside function. It represents the mass of all Lagrangian particles with values of the $p-$tuple greater than $(s_1,\cdots,s_n)$.\footnote{ 
\citet{MethvenB15} used a similar approach to construct a background state in an atmospheric context. 
}  
If we wish to express the theory in Eulerian coordinates, we just need to replace Eulerian coordinates $x^i=x^i({\bm \alpha},t)$ in (\ref{eq:C_int}) and integrate over the corresponding Eulerian mass density form
\begin{equation}
\mathfrak{D}({\bm x})=\left(\rho({\bm \alpha})\frac{\partial(\alpha^1,\cdots,\alpha^n)}{\partial(x^1,\cdots,x^n)}(t)\right)\mathfrak{V}({\bm x}),
\end{equation}
where $\mathfrak{V}({\bm x})$ is the volume form in Eulerian coordinates and $\rho({\bm \alpha})$ the scalar mass density in Lagrangian coordinates. 
By construction, the MDF  is a $p-$parameter family of Casimirs, one for each $p-$tuple.  MDFs introduce a phase space "foliation": with the holonomic brake turned off, $q(t)$ moves on a leaf of the foliation along isolines $H[q]$; when the brake is turned on, $q(t)$ still moves on the same leaf, but crosses the isolines of $H[q]$ on its way to $q_*$, which, by our hypothesis, is where the naive Hamiltonian attains its minimum on the leaf (see figure~\ref{fig:1}).
Two remarks are in order: While it may {be} possible to use (\ref{eq:C_diff}) to quantify the constraints, the integral version (\ref{eq:C_int}) has the advantage of including bulk constraints, such as conservation of total mass; also, for incompressible flows in the Boussinesq approximation, we can replace the mass $n-$form with the volume $n-$form, in which case we should properly speak of a Volume Distribution Function. For simplicity, we will still refer to $M(\bm s)$ as the MDF regardless of what $n-$form is used. 

To solve the minimization problem, we introduce  a suitable Lagrange multiplier $\psi({\bm s})$, defined on the support of the MDF $M^0(\bm s)$ that specifies the leaf. That is, 
we seek 
{to render the Lagrange} functional
\begin{equation}
L[q,\psi]\equiv H[q]+\int\psi({\bm s})\left[\int_{\mathbb{D}}\prod_{i=1}^p\theta(s_i({\bf x})-s_i)\mathfrak{D}({\bm x})-M^0({\bm s})\right]d{\bm s} \label{eq:E1}
\end{equation}
{stationary, }
by varying both $q$ and $\psi$ (recall that $s_i(\mathbf{x})$ is the value of the $i$-th Lagrangian invariant at $\mathbf{x}$ calculated from the fields in $q$), that is we seek the solution\footnote{Though we should admit the possibility that more than one extremal (minimal) solution exists, for simplicity we assume that the minimal solution is unique.} $q_*,\psi_*$  of
\begin{equation}
\frac{\delta L}{\delta q}\cong 0,\,\frac{\delta L}{\delta \psi}=0.\label{eq:E1.5}
\end{equation}
 If all the $s_i$'s depend algebraically on the field in $q$, then the $\cong 0$ can be replaced by equality {\em stricto sensu}.  
From now on, $q_*$ will be referred to as the "ground state". 
\subsection{The gauge-fixing condition}
Let 
\begin{equation}
\Psi[q]\equiv \int_{\mathbb{D}}\left[\int\psi({\bm s})\left(\prod_{i=1}^p\theta(s_i({\bf x})-s_i)\right)d{\bm s}\right]\mathfrak{D}(\bm x),\label{eq:E3}
\end{equation}
be the Casimir associated to the Lagrange multiplier and let
\begin{equation}
{\cal H}[q]=H[q]+\Psi[q].
\end{equation}
By switching the integral over the $p-$tuples $\bm s$ with the integral over the manifold $\mathbb{D}$, the Lagrange functional can be rewritten as 
\begin{equation}
L[q,\psi]=
{\cal H}[q]-\int\psi({\bm s})M^0({\bm s})d{\bm s}.
\label{eq:E2}
\end{equation}

Since the last term in (\ref{eq:E2}) does not depend on $q$, the Lagrangian multiplier $\psi_*$ associated to the ground state $q_*$ 
can be used to define the following  gauge-fixing condition to the equivalence class of Hamiltonians 
\begin{equation}
\left.\frac{\delta {\cal H}}{\delta q} \right|_{q_*}\cong  0,\label{eq:E4}
\end{equation}
which is satisfied when ${\cal H}[q]={\cal H}_*[q]\equiv H[q]+\Psi_*[q]$, where the Casimir $\Psi_*[q]$ is related to the Lagrange multiplier $\psi_*$ by (\ref{eq:E3}). 
We leave to the reader to verify that if two states $q_1$ and $q_2$ belong to the same leaf (i.e., they have the same MDF), then
\begin{equation}
\Psi_*[q_1]=\Psi_*[q_2].\label{eq:E5}
\end{equation}
We are now in a position to define the local diagnostic energy as follows: 
\begin{definition}
The local diagnostic energy $\cal E$ is the scalar density of the Hamiltonian which satisfies  the gauge-fixing condition (\ref{eq:E4}), with the ground state $q_*$ obtained from (\ref{eq:E1.5}) ,  i.e.
\begin{equation}
{\cal E}(q)=H(q)+\Psi_*(q).
\end{equation}
\end{definition}

In terms of the local diagnostic energy, 
the Available Energy  is 
\begin{equation}
E_{AE}[q]\equiv {\cal E}[q]-{\cal E}[q_*]=H[q]-H[q_*],
\end{equation}
where the last equality follows from (\ref{eq:E5}), since $q$ and $q_*$ belong to the same leaf. Thus, the gauge-fixing condition selects, out of the class of dynamically equivalent Hamiltonians, the one that returns the same Available Energy as the naive Hamiltonian, when referenced to the ground state. Of course, the naive Hamiltonian is physically interpreted as  being the energy of the system.     
We also make the following conjecture: There exists a neighborhood of $q_*$ over which  ${\cal E}[q]$ is a convex function (we implicitly assume that the phase space has a vector space structure). 
Note that (\ref{eq:E4}) is a necessary condition for the conjecture to hold. 
What makes the local diagnostic energy interesting is how the evolution of the Available Energy under full diabatic conditions is related to the local diagnostic energy.
\subsection{Evolution of the Available Energy under diabatic conditions}
The Hamiltonian selected by the gauge-fixing condition (\ref{eq:E4}) acquires a special role when we apply  non-holonomic forces  which break the  {invariance} of the $s_i$'s along Lagrangian trajectories. Without brakes, the motion occurs on a leaf along lines of constant energy. With a holonomic brake, the motion still occurs within the leaf, crossing the isolines if the naive Hamiltonian. Finally with the full non-holonomic brake on, the MDF becomes time dependent and  $q(t)$ drifts across leaves. At each time, we can still calculate a ground state solving (\ref{eq:E1}-\ref{eq:E1.5}) with $M^0({\bm s})$ replaced by its instantaneous value $M^t({\bm s})$  calculated from  $q(t)$. Thus, starting from a time series of $q(t)$ states, we obtain a sequence of ground states $q_*(t)$ with the corresponding Lagrange multipliers $\psi_*({\bm s},t)$. Note that (\ref{eq:E5}) still holds on the pairs $(q(t),q_*(t))$.

Based on our definition of local diagnostic energy, 
\begin{equation}
{\cal E}[q_*(t)]=H[q_*(t)]+\Psi_*[q_*(t),t],
\end{equation}
where the second argument in $\Psi_*[\cdot,t]$ is a reminder that the time dependence of the Casimir is due both to the time dependence of the ground state as well as to the intrinsic time dependence of $\psi_*$. The rate of change of the integrated local diagnostic energy of the ground state is 
\begin{equation}
\frac{\mathrm{d}{\cal E}[q_*(t)]}{\mathrm{d}t}=\left. \left(\frac{\delta H}{\delta q}+\frac{\delta \Psi_*}{\delta q}\right)\right|_{(q_*(t),t)}\left[\frac{\mathrm{d} q_*}{\mathrm{d}t}\right]+\left.\frac{\partial\Psi_*}{\partial t}\right|_{(q_*(t),t)}\cong \left.\frac{\partial\Psi_*}{\partial t}\right|_{(q_*(t),t)},
\end{equation}
where the last equality is made possible by the gauge-fixing condition.
Note that the last partial derivative only applies to the {\em explicit} dependence on time in the gauge-fixed Casimir\footnote{ In the following, when we speak of {\em the} Casimir, we will intend the gauge-fixed Casimir, as opposed to {\em a} generic Casimir. The context will usually suffice to make the distinction clear.  }. Since $q(t)$ and $q_*(t)$ are on the same leaf, we have 
\begin{equation}
\frac{\mathrm{d}{\cal E}[q_*(t)]}{\mathrm{d}t}\cong \left.\frac{\partial\Psi_*}{\partial t}\right|_{(q(t),t)},
\end{equation}
where now the Casimir is evaluated on  $q(t)$, rather than on the ground state $q_*(t)$. 
Therefore, the rate of change of the Available Energy 
\begin{equation}
\frac{\mathrm{d}E_{AE}[q(t)]}{\mathrm{d}t}\cong\left.\frac{\delta{\cal E}}{\delta q}\right|_{(q(t),t)}\left[\frac{\mathrm{d}q}{\mathrm{d}t}\right]\cong \left. \left(\frac{\delta H}{\delta q}+\frac{\delta \Psi_*}{\delta q}\right)\right|_{(q(t),t)}\left[\frac{\mathrm{d} q}{\mathrm{d}t}\right],
\end{equation}
can be expressed solely in terms of the rate of change of the local diagnostic energy evaluated on $q$. 
The rate of change of the naive Hamiltonian reflects the action of the holonomic and non-holonomic components of the diabatic processes, while the change in the Casimir is due solely to the non-holonomic component. Interestingly, the non-holonomic component may actually increase the Available Energy, at least temporarily and/or locally. 
 The geometric structure of the leaf provides the "maximal" constraint:  the evolution of the system in the presence of non-holonomic forces can relax the constraint, and in so doing  frees up energy that would have been frozen in the ground state. 
\section{Specific fluid models}\label{sec:fluids_models}
 The theory developed in the preceding sections is quite general, and applies to any type of  flow regime that can be described in the adiabatic limit with a Hamiltonian structure having a degenerate Poisson algebra. 
In this section, we apply the theory to two types of flows: Shallow water flows, and 
continuously stratified flows that can be studied in the Boussinesq approximation. 
The first is chosen because it is often used as a first approximation to describe large-scale geophysical flows. Since it has only one Lagrangian invariant, it provides a good pedagogical introduction.  The second is commonly used to study low-Mach number flows in which vertical oscillations are confined to vertical scales much smaller than a suitably defined compressibility scale.  
While there are issues with the Boussinesq approximation, especially regarding the proper relation to  thermodynamics \citep{Tailleux13}, it nonetheless provides a good model to study flows in which the internal energy reservoir, in an approximate sense, merely acts as a sink of energy dissipated via diabatic effects. 
Most oceanographic models are based on the Boussinesq approximation, and therefore it is worthwhile to provide energy-based diagnostic tools applicable to Boussinesq flows.

The main result is that in the flows considered here,  the Casimir is related via a simple transformation to the Bernoulli invariant  of the ground state, and that in strongly rotating systems, to leading order, the Casimir is independent on the details of the MDF. These results can be obtained relatively easily in the shallow water case. The generalization to Boussinesq flows is more involved. For this reason, we will isolate the main results as theorems and corollaries to allow readers who are not interested in the derivation  to skip over the mathematical details.   
\subsection{Shallow water flows}
Shallow water flows provide a relatively simple system where we can apply  the ideas outlined in the previous section. While relevant in its own from a Geophysical Fluid Dynamics point of view, it is mostly used as a  propaedeutic tool, having the advantage of possessing a single Lagrangian invariant, the potential vorticity. We will keep the discussion informal. 
A more formal approach will be reserved for the more interesting case of continuously stratified flows in the Boussinesq approximation. 

Shallow water flows are described by the pair $(h,\lambda)$, where $h$ is the local surface elevation, measured relative to a surface of constant geopotential height\footnote{In a rotating system, the geopotential height includes the "potential" of the centrifugal force.},  and,   $\lambda=\bm v^\flat=v_idx^i$ the  $1$-form associated via the $\flat$ musical isomorphism to the velocity vector ${\bm v}=v^i{\bm\partial}/{\bm {\partial x}}^i$.  Here and thereafter, the Einstein convention of summation over repeated indexes is applied to Latin indexes.  Together, the pair $(h,\lambda)$  satisfies 
\begin{subequations}
\begin{eqnarray}
\frac{\partial\lambda}{\partial t}=-i_{\bm v}\Omega-d(gh+E_k),\label{eq:SW0a}\\
\frac{\partial h\mathfrak{A}}{\partial t}+{\cal L}_{\bm v}(h\mathfrak{A})=0,\label{eq:SW0b}
\end{eqnarray}
\end{subequations}
where  $\mathfrak{A}$  is the area element (a $2$-form) of the manifold which contains the flow, $E_k$ is the kinetic energy,  and
 the total vorticity $\Omega=d\lambda+\Xi$ is  the sum of the relative vorticity $d\lambda$ and the frame vorticity $\Xi$. In an inertial frame $\Xi=0$.  . Because of (\ref{eq:SW0b}), the role of $\mathfrak{D}$ is played by $h\mathfrak{A}$. 
For a flow in the shallow water approximation, the only Lagrangian invariant (up to a constant multiplicative factor) is the potential vorticity
\begin{equation}
p\equiv \frac{\omega}{h},
\end{equation}
where $\omega=\star\Omega$ is the Hodge-dual of the  vorticity $\Omega$.  

The naive Hamiltonian  is
\begin{equation}
{ H}[q]=\int_{\mathbb{D}} h\left(E_k+g\frac{h}{2}\right)\mathfrak{A}.
\end{equation}

The Casimir that satisfies the gauge fixing condition takes the form
\begin{equation}
\Psi[q]=\int_{\mathbb{D}} h\Psi(p){\mathfrak{A}}.
\end{equation}
A holonomic brake can be obtained adding to the r.h.s. of (\ref{eq:SW0a}) a term proportional to $d(\star d(h\Phi))$. Physically, it represents a conservative (so as not to affect the vorticity) force field that extracts energy from surface waves.  
\subsubsection{Ground state and gauge-fixed Casimir}
The ground state on a leaf characterized by the MDF $M(s)$, and the Casimir that specifies the gauge-fixed Hamiltonian are found solving
\begin{subequations}
\begin{eqnarray}
\frac{\delta{\cal H}}{\delta h}=(E_k+g h+\Psi-p\Psi_{,p})\mathfrak{A}=0,\label{eq:SW1a}\\
\frac{\delta{\cal H}}{\delta \lambda}\cong h(\star\lambda)+d\Psi_{,p}=0,\label{eq:SW1b}\\
\int_\mathbb{D} h\theta(p-s)\mathfrak{A}=M(s),\,p=\frac{\star
(d\lambda+\Xi)}{h}.\label{eq:SW1c}
\end{eqnarray}
\end{subequations}
for the three unknown  $h_*$, $\lambda_*$ and $\Psi_*(p)$.    To avoid notational clutter, we denote with $(\cdot)_{,s_1\ldots s_p}$ the $p$-th derivative w.r.t. $s_1,\ldots,s_p$ (note that the indexes are to the right of the comma). We assume that $M$ is differentiable, and that $M_{,s}<0$. In other words, we assume that there is a continuous distribution of potential vorticity, so that with the introduction of a suitable second coordinate $t$, $(p,t)$ can be used as local\footnote{It is important to realize that more than one chart may be needed to cover the entire manifold $\mathbb{D}$.} coordinates over the manifold.

Before considering actual solutions to (\ref{eq:SW1a}--\ref{eq:SW1c}), 
we point out a few general features. 
\begin{enumerate}
\item Eq.~(\ref{eq:SW1a}-\ref{eq:SW1c}) is a system of three equations relating 4 quantities $(\lambda_*,h_*,\Psi_*,M)$. The first two equations define a steady state solution of the shallow water equations in terms of $\Psi_*$. 
Indeed, from (\ref{eq:SW1b}) we have, with $\Phi_*\equiv-\star\lambda_*$ being the flux form, 
\begin{equation}
d(h_*\Phi_*)=d(d(\Psi_{*,p}))=0,
\end{equation}
by Poincare's lemma, thus (\ref{eq:SW0b}) is satisfied. 
Substituting $d\lambda_*=-\star ph_*=-ph_*\mathfrak{A}$ in the l.h.s. of (\ref{eq:SW0a}) and using (\ref{eq:SW1a}) we have
\begin{equation}
-i_{\bm v}(ph_*\mathfrak{A})+d(p\Psi_{*,p}-\Psi_*)=-ph_*\Phi_*+p\Psi_{*,pp}dp=-pd(\Psi_{*,p})+p\Psi_{*,pp}dp=0,
\end{equation}
thus satisfying (\ref{eq:SW0a}).

\item If $\Psi_*$ is a solution, so is $\Psi_*+ap$, where $a$ is any constant. This indeterminacy simply reflects the fact that to fully determine the solution we need to specify the circulation on the boundary of the manifold. Indeed, the Casimir associated to $ap$ is
\begin{equation}
a\int_{\mathbb{D}} hp\mathfrak{A}=a\int_\mathbb{D} d\lambda=a\int_{\partial\mathbb{D}}\lambda=a\Gamma, 
\end{equation}
where $\Gamma$ is the circulation along the boundary of the manifold. In particular, if the manifold is closed (i.e., has no boundary) $\Gamma=0$. 
\item The first equation tells us that the Casimir is related to the quantity  $E_{*k}+gh_*$  familiar from Bernoulli's theorem. Though it does not appear to have an agreed-upon name, for convenience we will refer to it as the Bernoulli invariant. 
Thus, if we know $h_*$ and $\lambda_*$, say by integrating (\ref{eq:SW0a}-\ref{eq:SW0b}) with a holonomic brake until a steady state is reached, we obtain  the Casimir via 
\begin{equation}
\Psi_*=p\int_{p_{\rm min}}^p\frac{E_{*k}(s)+gh_*(s)}{s^2}ds.
\end{equation}
\end{enumerate}
\subsubsection{The natural coordinates of the ground state}
The form of the equations, and the considerations just made, suggest  that the ground state is more conveniently expressed with a "natural" set of coordinates, which in the present case are the potential vorticity and a second, time-like coordinate $t$. The latter is defined so that the relationship between "geometric" $(x^1,x^2)$ and natural coordinates $(p,t)$ is locally given integrating along the trajectories
\begin{equation}
dx^i=v_*^idt,\label{eq:B3}
\end{equation}
to which we associate the space-time $1-$forms $\phi^i\equiv dx^i-v_*^idt$ (the subscript $_*$ reminds us that these are ground state fields). 
To calculate the volume element in natural coordinates, consider the $2-$form defined in space-time 
\begin{equation}
\mathfrak{Q}=\sqrt{g}\phi^1\wedge\phi^2=\mathfrak{A}-\Phi\wedge dt.
\end{equation}
Here $\sqrt{g}$ is the square root of the determinant of the metric in geometric coordinates. Let $(p,t)\overset{\zeta}{\rightarrow} (x^1,x^2)$ be the (local) map from natural to geometric coordinates, and let $(p,t)\overset{\hat\zeta}{\rightarrow}(\zeta(p,t),t)$ be the map that embeds space into space-time. Under the pullback $\hat\zeta^*\mathfrak{Q}=0$ or
 \begin{equation}
  \mathfrak{A}=\Phi\wedge dt=\frac{\Psi_{,pp}}{h}dp\wedge dt,\label{eq:B4}
 \end{equation}
which gives the volume form in natural coordinates. Alternatively, we can define $t$ as the second coordinate such that the r.h.s. of (\ref{eq:B4}) is the volume form of the manifold  (from now on, when multiplying simple differentials, we will follow standard practice and omit the exterior product symbol $\wedge$). 
Incidentally, note that (\ref{eq:B3}) is integrable since 
\begin{equation}
d\phi^i\wedge\mathfrak{Q}=0,\,i=1,2.
\end{equation}

In natural coordinates, $\bm v=\bm\partial/\bm \partial\bm t$, thus if $\Omega=\omega dpdt$ is the vorticity, from (\ref{eq:SW0a}) we have the following integrability condition
\begin{equation}
0=d(i_{\bm v}\Omega_*)=-d(\omega_* dp)=\omega_{*,t} dpdt.\label{eq:SW1.3} 
\end{equation}

We want to consider under what conditions the ground state coincides with a constant geopotential surface, i.e. when $dh=0$ (recall that $h$ is measured relative to the geopotential). Since the total volume is conserved, such a ground state has the minimum attainable potential energy consistent with conservation of volume. Setting $h_*=\mathrm{const.}$  in (\ref{eq:SW0a}) we obtain
\begin{equation}
0=\omega_* dp-dEk_*.\label{eq:SW1.31} 
\end{equation}
The MDF is frame-invariant, by which we mean that observers in different frames of reference (inertial or otherwise) will measure the same MDF. 
Thus, we can always take the point of view of an  
inertial observer, for which the vorticity is given by 
\begin{equation}
d\lambda=d(\star\Phi)=d(g_{pt}dp+g_{tt}dt)=(g_{tt,p}-g_{pt,t})dpdt,
\end{equation}
where the $g_{XY}=g_{ij}x^i_{,X}x^j_{,Y}$  are the covariant components of the metric in natural coordinates.
Given that $g_{tt}=g_{ij}x^i_{,t}x^j_{,t}=2E_{k*} $, we have 
\begin{equation}
\omega_*=Ek_{*,p}-x^i_{,p}g_{ij}\left(x^j_{,tt}+\Gamma^j_{kn}x^k_{,t}x^n_{,t}\right),
\end{equation}
where the $\Gamma^j_{km}$'s are the Christoffel symbols describing the Levi-Civita connection of the manifold $\mathbb{D}$. Substituted in (\ref{eq:SW1.31}) we obtain the following two conditions
\begin{subequations}
\begin{eqnarray}
E_{k*,t}=0,\label{eq:SW1.32a}\\
x^i_{,p}g_{ij}\left(x^j_{,tt}+\Gamma^j_{kn}x^k_{,t}x^n_{,t}\right)=0.\label{eq:SW1.32b}
\end{eqnarray}
\end{subequations}
The first condition states that the kinetic energy must be constant along streamlines. A sufficient condition for the second condition to be satisfied is that the streamlines are geodesics. 

\subsubsection{The low-Rossby number ground state of shallows flows in rotating channels}
A particularly interesting application of the observations made so far is to the case of a shallow water flow contained in a rotating channel, with Coriolis frequency $f$.  We seek a geostrophic ground state where the contribution of the relative vorticity $d\lambda$ to the total vorticity is $O({\rm Ro})$ relative the frame vorticity $\Xi$, the latter in Cartesian coordinates being $fdx^1dx^2$, where ${\rm Ro}$ is the Rossby number.
In a geostrophic ground state, the kinetic energy must be $O({\rm Ro}^2)$, and thus to leading order the Bernoulli invariant  is simply $gh_*(p)={gf}{p}^{-1}+O({\rm Ro})$. 
The MDF of such a ground state is narrowly distributed around a non-zero $\overline{p}$, i.e. 
\begin{equation}
M(p)=\overline{h}L_1L_2m\left(\frac{p-\overline{p}}{\Delta p}\right).
\end{equation}
with ${\rm Ro}\equiv \Delta p/\overline{p}$ and $\overline{h}$ an average depth. Vice versa, it is possible to show that if the MDF is narrowly distributed, there is a rotating frame in which  the ground state is geostrophic. The interesting aspect of geostrophic ground states is that to leading order the Casimir is given by 
\begin{equation}
\Psi_*=-\frac{fg}{2p}+O({\rm Ro}),\label{eq:SW1.33}
\end{equation}
and thus to $O({\rm Ro})$  the local diagnostic energy does not depend on the details of the MDF. 
\subsubsection{Non-holonomic dynamics on geostrophic leaves}
The evolution of the Casimir of shallow-water geostrophic leaves under the influence of non-holonomic forces provides an interesting perspective on how the erosion of the potential vorticity constraint affects the Available Energy of a system.  For simplicity, we choose a simple model for the non-holonomic dynamics, which includes the effects of the bottom frictional Ekman layers on the vorticity dynamics \citep[][p. 215]{Pedlosky86}. Denoting with $T$ the spin-down time of an Ekman layer, the potential vorticity satisfies 
\begin{equation}
\frac{\partial p}{\partial t}+{\cal L}_{\bm v}(p)=-\frac{\star d\lambda}{Th},
\end{equation}
i.e., the potential vorticity changes along trajectories by an amount proportional to the relative vorticity scaled with the local depth (Note that in a geostrophic flow, the r.h.s. is $O({\rm Ro\,\overline{p}T^{-1}})$.). Thus, the change of the Casimir density along trajectories is to leading order
\begin{equation}
\frac{\partial h\Psi_*}{\partial t}\mathfrak{A}+{\cal L}_{\bm v}(h\Psi_*\mathfrak{A})=\Psi_{*,p}\left(\frac{\partial p}{\partial t}+{\cal L}_{\bm v}(p)\right)=-\frac{fg}{2p^2}\frac{\star d\lambda}{T}\mathfrak{A}.
\end{equation}
Therefore, Ekman layers with cyclonic relative vorticity  act as a sink of Available Energy along trajectories, whereas layers with anticyclonic relative vorticity are a source of Available  Energy. Of course, both cyclonic and anticyclonic layers act as a net sink on the naive energy.  
\subsection{Stratified Boussinesq flows}
A continuously stratified flow in the Boussinesq approximation is described by the pair $(b,\lambda)$, where $b\equiv g(\rho_0-\rho)/\rho_0$ is the buoyancy,  and, as before,  $\lambda\equiv {\bm v}^\flat=v_idx^i$ satisfying 
\begin{subequations}
\begin{eqnarray}
\frac{\partial\lambda}{\partial t}=-i_{\bm v}\Omega-d(P+E_k)+bdZ,\label{eq:B0a}\\
\frac{\partial b}{\partial t}+i_{\bm v}db=0,\label{eq:B0b}\\
d\Phi=0,\label{eq:B0c}
\end{eqnarray}
\end{subequations}
where $P$ is the pressure, $\Phi=\star\lambda=i_{\bm v}\mathfrak{V}$ is the flux form, and $Z$ the geopotential height \citep{Holm08}. 
As before, the total vorticity $\Omega\equiv d\lambda+\Xi$ is the sum of the relative vorticity $d\lambda$ and the frame vorticity $\Xi$. Also, we assume that the geometry of the manifold $\mathbb{D}$ that contains the fluid is Euclidean. Thus, there exists geometric coordinates $(x^1,x^2,x^3)$ where the metric tensor assumes the simple form $g_{ij}=1$ if $i=j$ and $g_{ij}=0$ if $i\neq j$.   
Also note that (\ref{eq:B0c}) implicitly assumes that the volume form $\mathfrak{V}$ is constant in time. While the extension to non-inertial frames whose volume form is not constant in time is possible, in the interest of simplicity we do not pursue here. 

In a continuously stratified Boussinesq fluid there are two independent Lagrangian invariants, the buoyancy $b$ itself and the potential vorticity $p\equiv\star (db\wedge\Omega)$.
Note that if $f(b)$ is any (smooth) function of $b$, then $f(b)$ is also a Lagrangian invariant, and so is $\star(df(b)\wedge\Omega)$. This will be reflected in certain degrees of freedom available in the definition of the gauge-fixed Casimir.  
The mass form $\mathfrak{D}$ can be replaced by the volume form $\mathfrak{
V}$ of the manifold that contains the flow.  
 A holonomic brake  is provided by adding a term proportional to $(i_{\bf v}db)db$ to the r.h.s. of (\ref{eq:B0a}), which extracts energy from the internal wave field without affecting the component of the vorticity aligned along the buoyancy gradient. 
 
 The naive Hamiltonian is 
 \begin{equation}
 H[q]=\int_{\mathbb{D}}(E_k-bZ)\mathfrak{V},\label{naiveHam}
 \end{equation}
 where $E_k$ is the kinetic energy per unit mass,
 while Casimirs  have the general form 
 \begin{equation}
 \Psi[q]=\int_{\mathbb{D}}\Psi(b,p)\mathfrak{V}.\label{eq:Cas}
 \end{equation}
 In terms of the Lagrange multiplier, the density of the gauge-fixed Casimir is 
 \begin{equation}
  \Psi(b,p)=\int_{b_{\rm min}}^b\left(\int_{p_{\rm min}}^p\psi(q,s)dq\right)ds.
 \end{equation}
 
 \subsubsection{The gauge-fixing condition}
 The Fr\`echet derivatives of the naive energy w.r.t. to $\lambda$ and $b$ are trivial. We concentrate here on the variations of the Casimir written as in (\ref{eq:Cas}). 
 The variation w.r.t. to $b$ is given by
 \begin{equation}
 \begin{split}
 \frac{\delta\Psi}{\delta b}\delta b=(\Psi_{,b}\delta b+\Psi_{,p}\delta p)\mathfrak{V}=(\Psi_{,b}\delta b+\Psi_{,p}(\star(d(\delta b)\wedge\Omega)))\mathfrak{V}=\\
 \Psi_{,b}\delta b\mathfrak{V}+\Psi_{,p}d(\delta b)\wedge\Omega\cong \Psi_{,b}\delta b\mathfrak{V}-d(\Psi_{,p}\Omega)\delta b=\\ [(\Psi-p\Psi_{,p})_{,b}\mathfrak{V}-\Psi_{,pp}dp\wedge\Omega]\delta b\label{eq:B1a}
 \end{split}
  \end{equation}
  while the variation w.r.t. $\lambda$ is given by 
  \begin{equation}
  \frac{\delta\Psi}{\delta\lambda}\wedge \delta\lambda=\Psi_{,p}\delta p\mathfrak{V}=\Psi_{,p}db\wedge d(\delta\lambda)\cong d(\Psi_{,p}db)\wedge \delta\lambda.
  \end{equation}
 Given the MDF $V(p,b)$, the ground state and the gauge-fixed Casimir are found solving
 \begin{subequations}
 \begin{eqnarray}
 \frac{\delta\cal H}{\delta b}\cong[-Z+(\Psi-p\Psi_{,p})_{,b}]\mathfrak{V}-\Psi_{,pp}dp\wedge\Omega= 0\label{eq:B2a},\\
 \frac{\delta\cal H}{\delta\lambda}\cong\Phi+d(\Psi_{,p}db)= 0,\label{eq:B2b}\\
 \frac{\delta\cal H}{\delta\psi}=\int_\mathbb{D}[\theta(p(\bm x)-p)\theta(b(\bm x)-b)]\mathfrak{V}=V(p,b)\label{eq:B2c}
 \end{eqnarray}
 \end{subequations}
 An attentive reader may have wondered why we did not enforce the incompressibility condition (\ref{eq:B0c}) via an explicit Lagrangian multiplier. This is not necessary, as (\ref{eq:B2b}) guarantees that the flux form of the ground state is closed. Also, as before, we assume that the MDF is a continuous and differentiable function, i.e. buoyancy and potential vorticity are continuously distributed. More complicated cases consisting of layers where buoyancy and/or potential vorticity are interleaved with layers where one or the other are constant will not be considered here. 
 
 \subsubsection{The gauge-fixed Casimir} 
 The purpose of this section is to prove the following
 \begin{theorem}
Up to an inconsequential constant, the gauge-fixed Casimir is given by
\begin{equation}
\Psi_*=p\int_{ p_{\rm min}}^p\frac{B_*(s,b)}{s^2}ds,
\end{equation}
where $B_*=Ek_*(p,b)-bZ_*(p,b)+P_*(p,b)$ is the Bernoulli invariant 
of the ground state expressed in natural coordinates. 
\end{theorem}
As we did for the shallow water case, we introduce the natural set of coordinates, which now include  the buoyancy,  the potential vorticity and the time-like coordinate, in terms of which the volume 3-form is $\mathfrak{V}=\Psi_{,pp}\, dbdpdt$.  
In natural coordinates, we write the total vorticity as 
\begin{equation}
\Omega_*=\Omega^bdpdt+\Omega^pdtdb+\Omega^tdbdp. 
\end{equation}
Since $\Omega$ is closed by definition
\begin{equation}
(\Omega^b)_{,b}+(\Omega^p)_{,p}+(\Omega^t)_{,t}=0.
\end{equation}
From the definition of potential vorticity, 
\begin{equation}
p=\star(db\wedge\Omega)=\Omega^b\star(dbdpdt)=\Omega^b(\Psi_{,pp})^{-1}.\label{eq:B4.9}
\end{equation}

We introduce $\overline{\Psi}_*=p\Psi_{*,p}-\Psi_*$. 
Taking the Hodge star of (\ref{eq:B2a}) and using  (\ref{eq:B4}) we get
\begin{subequations}
\begin{eqnarray}
\overline{\Psi}_{*,b}=-(Z_*+\Omega^p).\label{eq:B5a}\\ 
\overline{\Psi}_{*,p}=\Omega^b,\label{eq:B5b}
\end{eqnarray}
\end{subequations}
where we have used the fact that  $\overline{\Psi}_{*,p}=p\Psi_{*,pp}$ and (\ref{eq:B4.9}). These two equations can be combined into a single equation 
\begin{equation}
d(\overline{\Psi}_*+Z_*b)=bdZ-\Omega^pdb+\Omega^bdp=bdZ_*-i_{{\bm v}_*}\Omega_*.\label{eq:B5.1}
\end{equation}
We have derived the above expression in natural coordinates, where ${\bm v}_*$ has the simple expression  ${\bm v}_*=\bm \partial/\bm\partial \bm t$, but 
(\ref{eq:B5.1}) is coordinate-free. Thus,  comparing (\ref{eq:B5.1}) with (\ref{eq:B0a}), we see immediately that $\overline{\Psi}_*$ is, up to a constant, the Bernoulli invariant of the ground state, and thus we have proven the theorem. \citep[For a proof using standard vector calculus see][]{Holm08}

 \subsubsection{The apedic theorem and its corollaries}
 Let us introduce a general 
  \begin{definition}
  An apedic\footnote{From the Greek  \bcode{a)/pedos}, meaning level, flat.} ground state is a ground state such that $b_*=b_*(Z)$. By extension, the leaf in phase space to which an apedic ground state belongs is called an apedic leaf.  
  \end{definition} 
  To characterize under what conditions a ground state is apedic we have the following 
  \begin{theorem}
(Apedic theorem) Let the absolute vorticity of a ground state be 
\begin{equation}\Omega_*=\Omega^tdbdp+\Omega^bdpdt+\Omega^pdtdb.\end{equation} The ground state is  apedic if and only if
\begin{subequations}
\begin{eqnarray}
\Omega^p_{,t}=0,\label{eq:B5.5a}\\
\Omega^t_{,t}=0\label{eq:B5.5b},
\end{eqnarray}
\end{subequations}
that is, the absolute circulation over loops embedded in surfaces of constant potential vorticity \textit{and} the circulation over loops embedded in surfaces of constant "time" must be independent on "time" for a ground state to be apedic.
\end{theorem}
This follows from the integrability condition for (\ref{eq:B5.1}). Indeed, taking the external derivative of (\ref{eq:B5.1}) and leveraging the closedness of $\Omega$, we obtain after some simple algebra 
\begin{equation}
dZdb+\Omega^t_{,t}dbdp+\Omega^p_{,t}dtdb=0.\label{eq:B5.5}
\end{equation}
 The ground state must satisfy (\ref{eq:B5.5}). Since $dpdt$ and $dtdb$ are independent bases of the module of $2-$forms over $\mathbb{D}$, and  on an apedic manifold $dbdZ=0$ by definition, we have the proof of the apedic theorem.  

Let us now explore some consequences of the  apedic theorem. From (\ref{eq:B2b}) we know that in an inertial system, the vorticity of the ground state
\begin{equation}
\Omega_*=d\lambda_*=d\star(\Psi_{*,pp}dbdp)=d(g_{bt}db+g_{pt}dp+g_{tt}dt).\label{eq:B5.6}
\end{equation}
  For a manifold to be apedic, (\ref{eq:B5.6}) substituted in (\ref{eq:B5.5a}-\ref{eq:B5.5b}) requires that the metric satisfy  
\begin{subequations}
\begin{eqnarray}
g_{pt,bt}-g_{bt,pt}=0,\\
g_{bt,tt}-g_{tt,bt}=0.
\end{eqnarray}
\end{subequations}
In terms of Cartesian coordinates, the apedic condition becomes
\begin{subequations}
\begin{eqnarray}
x^i_{,p}x^i_{,btt}-x^i_{,b}x^i_{,ptt}=0,\label{eq:B6a}\\
x^i_{,b}x^i_{,ttt}-x^i_{,t}x^i_{,btt}=0\label{eq:B6b}.
\end{eqnarray}
\end{subequations}
which provides the proof to  the first apedic corollary
\begin{corollary}
A sufficient condition for a ground state to be apedic is that streamlines are geodesics and the velocity is uniform along streamlines.
\end{corollary}
Note that the Cartesian nature of the coordinates is of the essence. 
Consider, for example,  $x^i$'s that are cylindrical coordinates $(z,r,\phi)$ and a ground state flow such that in its natural coordinates the $x^i$'s at most depend linearly on $t$. Then, for example, we have from (\ref{eq:B5.5a})  that 
\begin{equation}
\Omega^t_{,t}=2\phi_{,t}r(r_{,b}\phi_{,pt}-r_{,p}\phi_{,bt}),
\end{equation}
which in general will not be equal to zero (and indeed, the trajectories are not geodesics!). In other words, the acceleration is relative to an inertial frame. Consequently, we can expect that when the system is non-inertial, a generic ground state will not be apedic. 
Also, it is worth remarking that the notion of apedicity, depends on the local direction of the plumb line, and thus it is not frame-independent. Thus, a fluid in solid body rotation will appear apedic to an observer in solid body rotation with the flow, but not to an inertial observer.

\subsubsection{The Casimir of inertial apedic ground states}\label{sec:apedic}
Under the assumption that the ground state is apedic in an inertial system, we can make further analytic inroads. Let $E_k=x^i_{,t}x^i_{,t}/2=g_{tt}/2$ be the kinetic energy of the ground state. Assuming that apedicity is due to the trajectories being inertial (recall that the first apedic corollary proves only the sufficiency condition), we have 
\begin{eqnarray}
\Omega^b=g_{tt,p}-g_{pt,t}=2Ek_{,p}-Ek_{,p}=Ek_{,p},\\
\Omega^p=-g_{tt,b}+g_{bt,t}=-2Ek_{,b}+Ek_{,b}=-Ek_{,b},
\end{eqnarray}
Then (\ref{eq:B5.5a}-\ref{eq:B5.5b}) can be combined
\begin{eqnarray}
d(\overline{\Psi}-Ek+Zb)=bdZ=d\left(\int^Zb\,\mathrm{d}Z\right),
\end{eqnarray}
and since we know that $\overline{\Psi}$ is the Bernoulli function of the ground state, we immediately deduce that the pressure distribution in apedic ground states is hydrostatic. 

The definitions of local Available Potential Energy proposed in the past that rely on a Casimir \citep[e.g.,][]{HollidayM81}, usually, \citep[but now always, see, e.g.][]{CodobanS03} include only the potential and pressure term, combined in
$\int^bZdb$. This is 
 not surprising, since 
the traditional Available Energy approach  is to define the Background Potential Energy as the energy of the isochorically restratified flow. In our formalism, it amounts to enforcing conservation of buoyancy but not potential vorticity, which is to say we are minimizing the energy on a wider leaf. If we want to enforce both buoyancy and potential vorticity conservation, the corresponding ground state in general must have kinetic energy, and thus the system will have a lower Available Energy. 
This, however, does not exclude the existence  of interesting classes of flows for which the traditional approach is valid. They will be discussed in the next section,   
where we apply the theory to simple channel geometries,  both inertial and rotating.

\subsubsection{Boussinesq Channel flows}
For  arbitrary MDFs and geometries, (\ref{eq:B2a}-\ref{eq:B2c}) constitute a formidable nonlinear problem for the ground state. In these situations, to calculate the Casimir it is better to solve by other means (more likely, numerically) the Euler equations with a holonomic brake to find the ground state. Once the pair $b_*,\lambda_*$ is known, the Casimir can be calculated from its Bernoulli invariant. 

Analytic progress can be made if we are willing to sacrifice geometric complexity. Thus, here we consider simple "open" channel geometries, which geometrically can be described by Cartesian coordinates. For definiteness, let $0\leq x^1\equiv Z\leq L_z$ be the vertical direction, $0\leq x^2\equiv y\leq L_y$ the spanwise direction and $0\leq x^3\equiv x\leq L_x$ the streamwise direction.  The bottom, top and side walls are free-slip boundaries, while periodicity in streamwise direction.  
We seek what we may call $2-1/2$ dimensional ground states: these are states such that $x^i=x^i(b,p),\,i=1,2$, and such that the velocity, which depends on two dimensions $(b,p)$ is normal to the $x^1,x^2$ plane (the 1/2 dimension).\\

\paragraph{Inertial channels and the second apedic corollary.} \label{sec:W}
In inertial channels, we have the following
\begin{corollary}
Let $V$ be the MDF of a flow contained in an inertial channel which can be described by a 2-1/2 dimensional ground state as defined above. Then the ground state is apedic. 
Further,  if, for any value of $b$, $V_{,pb}$ is an even function of $p$, the
kinetic energy of the ground state is zero.
\end{corollary}
The method to construct an apedic ground state for simple flow geometries extends the technique introduced by \cite{WintersLRD95} and consists of a series of steps:
\begin{itemize}
\item[--] Determine the resorted height $Z_{*}(b)$ solving (\ref{eq:B2c}) in the limit $p\to p_{\rm min}$. This coincides with the isochoric restratification of the fluid. 
\item[--] Compute the resorted distribution of potential vorticity along the spanwise direction $y_*(p)$.
\item[--] Finally, determine the kinetic energy $E_{k*}$ of the ground state.
\end{itemize}

Later, we will modify this recipe to calculate the ground state and associated Casimir of rotating flows characterized by a low-Rossby number ground state, with the Rossby number being defined as $Ro=O(\omega_*/f)$ where $\omega_*$ is the relative vorticity and $f$ is the Coriolis frequency.

It is straightforward to verify that $2-1/2$ dimensional ground states in inertial channels are apedic.  Indeed, in geometric coordinates, the components of the velocity vector are $[0,0,v_*(b(Z,y),p(Z,y))]$, thus $dx=v(b,p)dt$ and $dy=dZ=0$, hence the velocity is uniform along geodesics.  Rather than working in "natural" coordinates, we work with "hybrid" coordinates $(b,p,x)$. 
Being apedic $Z_*=Z(b)$, (here, as we have done before, quantities used as a coordinates will not carry the $_*$ subscript) which can be calculated from the limit $p\to p_{\rm min}$ of (\ref{eq:B2c})
\begin{equation}
\int\theta(b_*(Z)-s)dZdydx=-L_yL_x\int_{b_{\rm min}}^{b_{\rm Max}}\theta(b-s)\frac{dZ_*}{db}db=V(p_{\rm min},s),
\end{equation}
whose solution is 
\begin{equation}
Z_*(b)=L_z\left(1-\frac{V(p_{\rm min},b)}{V(p_{\rm min},b_{\rm min})}\right).\label{eq:B7}
\end{equation}
Its functional inverse $b_*(Z)$ is the buoyancy distribution of the ground state. From this, we can calculate the hydrostatic pressure of the ground state.  
Also, for later convenience, we define $N^{-2}_*\equiv Z_{*,b}(b)$. 

Next, consider the volume element
\begin{equation}
\mathfrak{V}=dZdydx=\left|\frac{\partial(Z,y)}{\partial(b,p)}\right|dbdpdx.
\end{equation}
Using again (\ref{eq:B2c}) 
\begin{equation}
\left|\frac{\partial(Z,y)}{\partial(b,p)}\right|=\frac{V_{,pb}}{L_x}, \label{eq:B7.1}
\end{equation}
from which $y=y(p,b)$ is obtained solving 
\begin{equation}
\frac{V_{,pb}}{L_x}=\left|\frac{\partial(Z,y)}{\partial(b,p)}\right|=\left|\frac{V_{,b}(p_{\rm min},b)}{L_xL_y}\frac{\partial y}{\partial p}\right|.
\end{equation}
The solution that spans the domain is
\begin{equation}
y_*=L_y\frac{V_{,b}(p_{\rm min},b)-V_{,b}(p,b)}{V_{,b}(p_{\rm min},b)}.
\end{equation}
Note that in this apedic ground state, isopycnals are flat, whereas surfaces of constant potential vorticity are 
not and that the relationship between natural and geometric coordinates is invariant under the dilation
$V\to e^s V$.

At this point, all we have left is to calculate the kinetic energy of the ground state. By construction, we must have
\begin{equation}
\lambda_*= Fdx,
\end{equation}
and since $\Omega=d\lambda=F_{,b}dbdx+F_{,p}dpdx$, using the definition of potential vorticity (\ref{eq:B4.9}) and (\ref{eq:B7.1}) we may try as a first \textit{Ansatz} a solution to 
\begin{equation}
F_{,p}=\frac{pV_{,pb}}{L_x},\label{eq:B7.15}
\end{equation}
 with the constant of integration set along planes of constant buoyancy so that the total momentum along the $x$ direction is conserved.

In the special case
\begin{equation}
V_{,pb}=\delta(p)\tilde{V}_{,b}, \label{eq:B7.2}
\end{equation}
which corresponds to a leaf characterized by a  uniformly zero distribution of potential vorticity,
$F$ and the kinetic energy of the ground state are zero.  The Casimir then reduces (up to a factor $pf(b)$ which does not concern us) to 
\begin{equation}
\Psi_*=\int_{b_{\rm min}}^bZ(s)ds,
\end{equation}
and the ground state to a quiescent fluid isochorically restratified according to (\ref{eq:B7}).
Two-dimensional flows contained in a vertical plane have zero potential vorticity, and thus  satisfy (\ref{eq:B7.2}).  In this case, we recover the same definition of Available Energy used by Winters {\it et al.}. 
\begin{figure}
\hspace{-1mm}
\hspace{-2mm}\includegraphics[scale=.46]{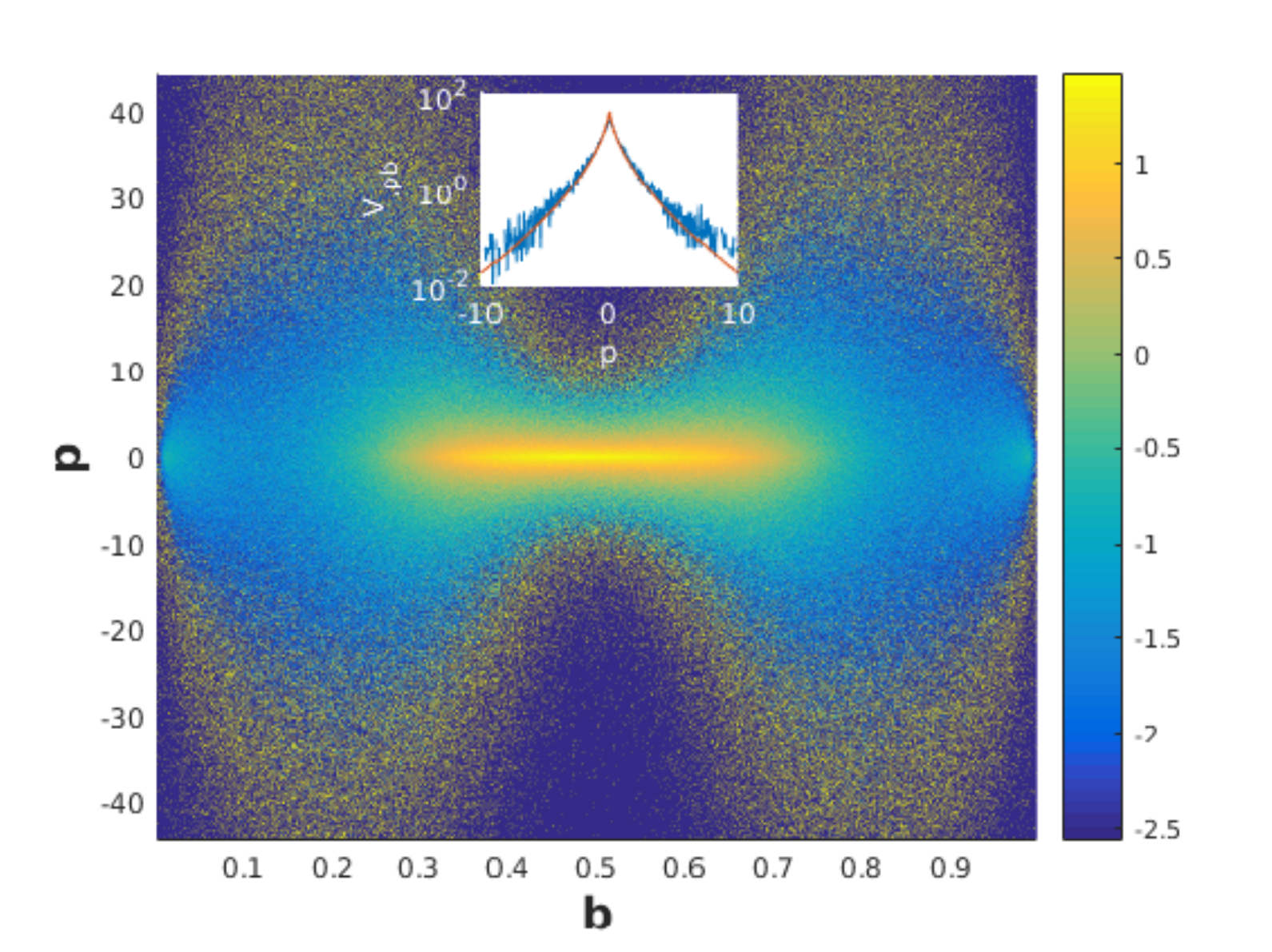}
\put(-205,0){(a)}
\hspace{-4mm}
\hspace{0mm}\includegraphics[scale=.46]{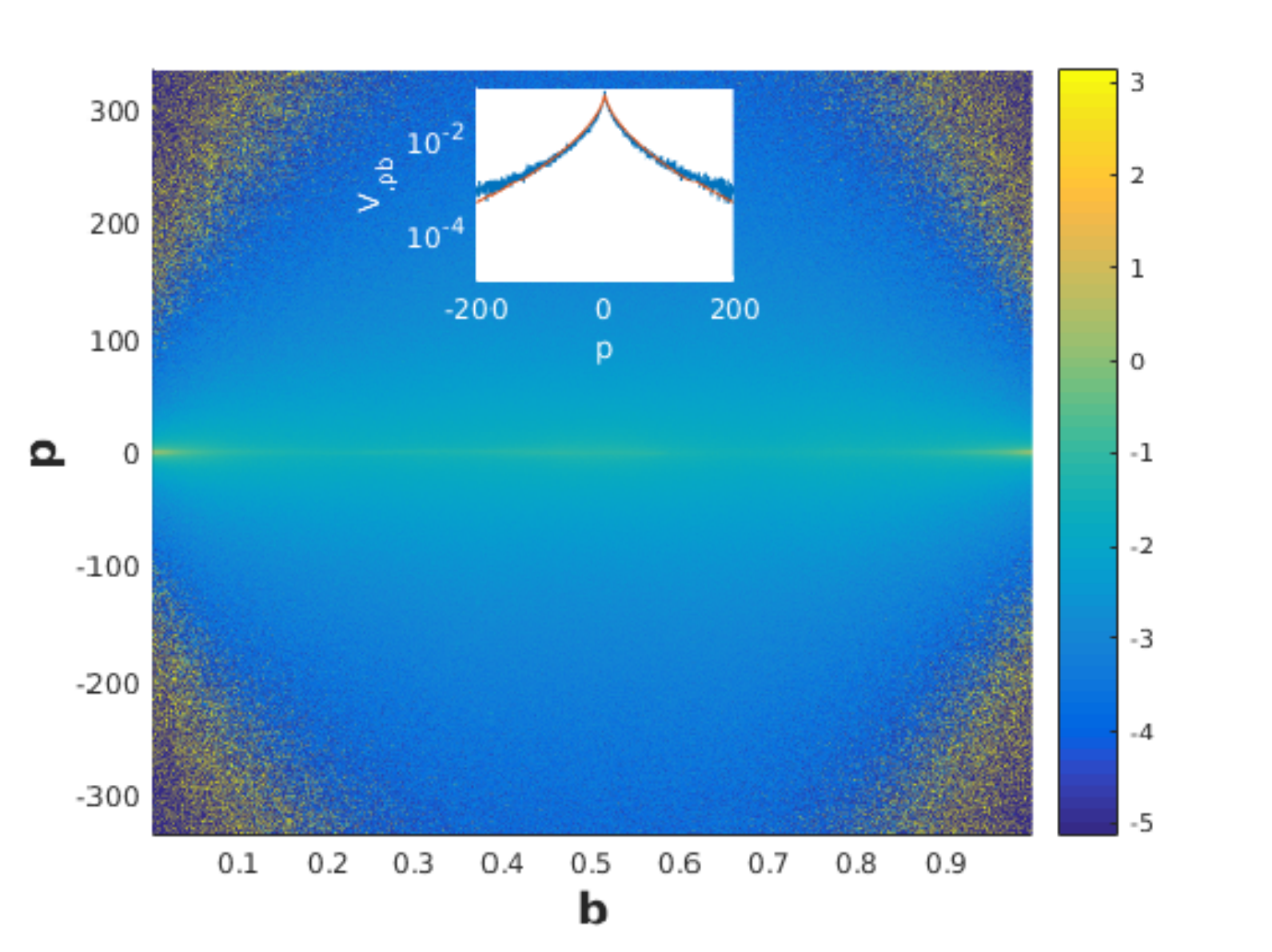}
\put(-210,0){(b)}
\caption{\label{fig:KH} $\log_{10}(V_{,pb})$ in stratified shear flows. (a) A stratified Couette flow: The inset shows $V_{,pb}(p,0.5)$ (blue line) superimposed to a stretched exponential $\exp{-|p/0.3|^{0.6}}$ (orange line). (b) A time evolving shear layer, after the initial break-up of the Kelvin-Helmholtz instabilities. In this case too, a stretched exponential with a similar exponent fits the profile ($\exp{-|p/4.5|^{0.6}}$). Note how, in both cases, profiles of $V_{bp}$ along lines of constant buoyancy are symmetric around the origin.} 
\end{figure}

\begin{figure}
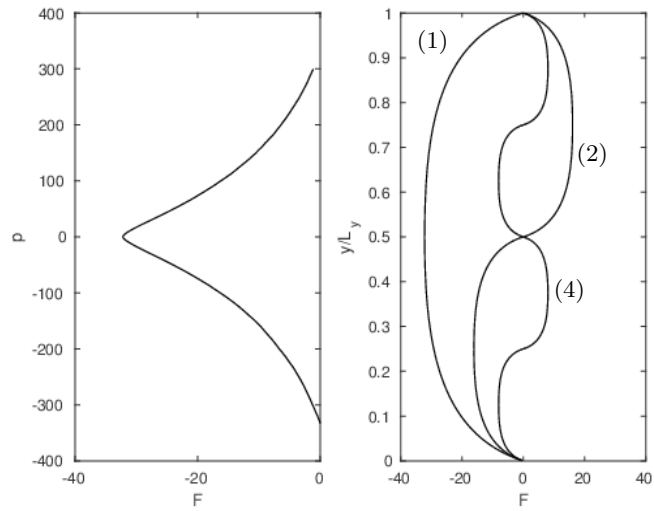

\begin{center}
\begin{lpic}{V_stitch(0.6)}
\lbl{100,105;$(1)$}
\lbl{135,80;$(2)$}
\lbl{130,50;$(4)$}
\end{lpic}
\end{center}
\caption{\label{fig:C1} Left side: $F(b_0,p)$ vs. $p$ where $F$ is obtained solving (\ref{eq:B7.15}) in a time-evolving stratified shear layer, where $b_0=(b_{\rm min}+b_{\rm Max})/2$ is the midplane buoyancy. Right side: $F(b_0,p)$ vs. $y(b_0,p)$ obtained using one (1), two (2) or four (4) charts to cover the domain.  }
\end{figure}

Consider a flow at time $t$ which evolved from a purely two-dimensional flow at $t=0$, i.e. at $t=0$ the MDF was $V_{,p}=\widetilde{V}_{,b}\delta(p)$. As instabilities develop, the delta-like distribution in the MDF develops tails. Let us assume that {in the absence of}  any mechanism that breaks the symmetry, the potential vorticity that the diabatic processes generate {\em ex nihilo}  {is} equally distributed around the origin. In other words, we assume that whatever mechanism generates potential vorticity, it is not biased against either positive or negative potential vorticity. 
Then, without loss of generality, we can write for the MDF at time $t$ as
\begin{equation}
V_{,pb}\simeq\frac{L_yL_xN_*^{-2}(b)}{\sigma(b)}w\left(\frac{p}{\sigma(b)}\right),\label{eq:B7.3}
\end{equation}
where $w(x)$ is a suitable even function. Analysis of numerical experiments indicate that a stretched exponential appears to give a good fit in different configurations (figure~\ref{fig:KH}). As long as it decays sufficiently rapidly, we can replace $p_{\rm min}$ with $-\infty$ and $p_{\rm Max}$ with $\infty$.  $\sigma(b)$ measures the spread {(i.e. the variance if the distribution was Gaussian)} in potential vorticity near a given buoyancy level. Denoting with $W(x)=\int_{-\infty}^x tw(t)dt$ and with $\overline{W}_n=\int_{-\infty}^{\infty}W^n(t)dt$, integration of (\ref{eq:B7.15}) yields for the velocity the following expression (setting the integration constant to zero for the moment) 
\begin{equation}
F(p,b)=\int_{-\infty}^pp\frac{V_{,pb}}{L_x}\mathrm{d}p=L_y W\left(\frac{p}{\sigma(b)}\right)N_*^{-2}(b)\sigma(b),
\end{equation}
from which the kinetic energy per unit volume  of the ground state can be easily calculated 
\begin{equation}
\begin{split}
Ek_*=\int_{-\infty}^{+\infty}\left(\int_{b_{\rm min}}^{b_{\rm Max}}\frac{F^2}{2}V_{,bp}\mathrm{d}b\right )\mathrm{d}p\left/\int_{-\infty}^{+\infty}\left(\int_{b_{\rm min}}^{b_{\rm Max}}V_{,bp}\mathrm{d}b\right )\mathrm{d}p\right.=\\
\frac{L_y^2}{2}\left(\frac{\overline{W}_2}{L_z}\int_0^{L_z}(N_*^{-2}(b_*(Z)\sigma(b_*(Z)))^2dZ\right),
\end{split}
\end{equation}
the inescapable consequence of which is that the kinetic energy {\em per unit volume} of this point on the leaf   grows with the width $L_y$ of the channel squared. While we have derived this result assuming that the MDF has the particular form given by (\ref{eq:B7.3}), it fundamentally rests on the fact that under the dilation $V\to e^{s}V$ we have $F\to e^{s} F$ and $Ek\to e^{2s} Ek$. Therefore, what we have derived cannot be the ground state that we are seeking. To understand what went wrong, and to find a cure for it, in figure~\ref{fig:C1} we plot $F$ calculated using (\ref{eq:B7.15}) where the MDF was obtained from a DNS of stratified Couette flow \citep{Scotti15} (indicated with the label 1).     
In this set up, the Lagrangian parcels with high negative potential vorticity are shunted on one side of the domain, most of the domain is filled by the (many) particles with little potential vorticity, while the parcels with large positive vorticity are pushed to the other side of the physical domain. We can see immediately two problems with this approach: First, having separated the large-negative   from the large-positive potential vorticity particles, we have created a large, sustained shear that forces the velocity to reach a large value in the center of the domain; and second, while this setup does not violate a free-slip boundary condition at the side walls, it cannot accommodate periodic boundary conditions. 

Indeed,  no matter what the details of the MDF are, a single $(b,p,x)\overset{\zeta}{\rightarrow}(z,y,x)$ chart cannot cover the manifold if the latter is periodic in the $y$ direction. 
Let us explore then what happens if we allow two charts to cover the manifold. The first chart maps the $(b,p,x)$ space to  the half of the channel such that $0\leq y\leq L_y/2$, while the second covers the other half. Further, since the MDF is an {\em extensive} quantity by definition, we require that each half of the physical domain contributes exactly one half to the total MDF. In each half, we proceed exactly as we did before, and since $F(p,b)\overset{p\to\pm\infty}\longrightarrow0$ it is possible to stitch together the velocity at the edge of every subdomain to obtain the profile shown on the right panel of figure \ref{fig:C1} labelled (2). Because in each half the MDF is half of the original one, the energy per unit volume in each half (an intensive quantity) is now 25\% of the energy per unit volume obtained using only one chart. As a bonus, the velocity field is now periodic. Again, what makes this possible is the fact that $V_{,pb}$ is an even function of $p$. 

Of course, there is no reason to stop at two charts. The procedure can be repeated with $4,8,\ldots$, charts, each covering $1/4,1/8,\ldots$ of the channel, and each time obtaining a field with a kinetic energy per unit volume $1/16,1/64,\ldots$ of the previous (figure~\ref{fig:C1} shows the profiles obtained using one, two and four charts). Thus, we generate a sequence of velocity fields in which potential vorticity is interleaved at finer and finer scales, while at the same time reducing the overall kinetic energy by a factor $4$  with each iteration. Barring the existence of a cut-off scale, we have proven the second apedic corollary.\\

Naturally, it is possible to consider initial conditions in which the symmetry in $p$ is broken {\em ab initio}, even in the absence of rotation. For example, consider   filling a channel, which, as before, has dimensions $(L_z,L_y,L_x)$ with a fluid whose  buoyancy 
has a uniform gradient along the spanwise direction $db=N_0^2dy$, and flowing such that $\lambda=\Delta v_0 (Z/L_z)dx$, so that $\Omega=-\Delta v_0/L_zdxdz$. Thus, the potential vorticity is uniform and equal to $-N^2_0\Delta v_0/L_z$. We can handle this singular limit by "smearing" the potential vorticity distribution around its constant value by an amount $O(\epsilon)$, carry out the calculation, and finally taking the $\epsilon\to0$ limit. In this case, the buoyancy of the ground state is such that   $db_*=N^2_0(L_y/L_z)dZ$ and $\lambda_*=\Delta v_0(y/L_y)dx$. Interestingly, the ground state has as much kinetic energy as the initial state, and  thus the Available Energy per unit volume is simply the difference between the potential energy of the initial and ground state $(\Delta b L_z/12)$. 


\paragraph{Rotating channels: geostrophic ground states.} 
We want to consider here the ground states of flows contained in rotating channels where the frame vorticity is much larger than the relative vorticity.   In other words, we look for ground states characterized by a small value of the Rossby number ${\rm Ro}\equiv O(\omega/f)$, where $\omega$ is the order of magnitude of the relative vorticity and $f$ the Coriolis frequency.  As we did in the shallow-water case, we  call low-${\rm Ro}$ ground states {\em geostrophic} ground states. We work with natural $(b,p,t)$ coordinates assuming that the ground state is 2-1/2 dimensional. 

Under these assumptions, the main source of non-inertiality along trajectories in the ground state is due to the frame vorticity, which we write as 
\begin{equation}
\Xi=fdydx=\Xi^tdbdp+\Xi^bdpdt+\Xi^pdtdb,\label{eq:B7.9}
\end{equation}
with
\begin{subequations}
\begin{eqnarray}
\Xi^t=f(y_{,b}x_{,p}-y_{,p}x_{,b}),\label{eq:B7.9a}\\
\Xi^b=fy_{,p}x_{,t},\label{eq:B7.9b}\\
\Xi^p=-fy_{,b}x_{,t},\label{eq:B7.9c}
\end{eqnarray}
\end{subequations}
where $f$ is the Coriolis parameter.

Under the 2-1/2 assumption,  $x_{,t}=\sqrt{2Ek}$, and (\ref{eq:B5.1}) becomes
\begin{equation}
d\overline{\Psi}=f\sqrt{2Ek}\,dy-Zdb+O({\rm Ro}),\label{eq:B8.1}
\end{equation}
where we made use of (\ref{eq:B7.9b}-\ref{eq:B7.9c}). 
To $O({\rm Ro})$ the integrability condition 
\begin{equation}
f\sqrt{2Ek}_{,Z}dZdy=b_{,y}dZdy,\label{eq:B8.2}
\end{equation}
shows that the ground state is in the so-called thermal wind balance \citep{Pedlosky86}. As we did for flows in inertial channels, we endeavor to construct 
the ground state explicitly by seeking first  the coordinate maps $Z_*=Z(b,p)$ and $y_*=y(b,p)$. 

Along a streamline extending from one end of the domain to the other, $\int dt=L_x/\sqrt{2Ek}$. From the definition of MDF
\begin{equation}
\frac{L_x}{\sqrt{2Ek}}\Psi_{,pp}= {V}_{,pb},
\end{equation}
which, combined with (\ref{eq:B5b}) gives
\begin{equation}
fy_{*,p}=\frac{p{V}_{,pb}}{L_x}.\label{eq:B9}
\end{equation}
A second equation involving the coordinates is (\ref{eq:B7.1}),
\begin{equation}
\left|\frac{\partial(Z_*,y_*)}{\partial(b,p)}\right|=Z_{*,b}y_{*,p}-Z_{*,p}y_{*,b}=\frac{{V}_{,pb}}{L_x}. \label{eq:B10}
\end{equation}
In choosing the sign, we assume that $f>0$. 
  We seek solutions of (\ref{eq:B9}-\ref{eq:B10}) in power series of $(b-b_{\rm min})$ as follows
  \begin{subequations}
  \begin{eqnarray}
  Z_*=\sum_{n=1}^\infty Z_n(p)(b-b_{\rm min})^n,\label{eq:B10a}\\ 
  y_*=\sum_{n=0}^\infty y_n(p)(b-b_{\rm min})^n,\label{eq:B10b}
  \end{eqnarray}
  \end{subequations}
 where we assume that the level of zero motion is the flat isopycnal $b=b_{\rm min}$. We also write
 \begin{equation}
 \frac{V_{,pb}}{L_x}=\sum_{n=0}^{\infty}F_n(p)(b-b_{\rm min})^n.
 \end{equation}
 as power series.\footnote{From a computational point of view, it may be more advantageous to use orthogonal polynomials. Here, we use a simple power series for convenience.} Inserting (\ref{eq:B10b}) into (\ref{eq:B9}) we obtain the coefficients of the power series for $y$ as 
 \begin{equation}
 y_n=f^{-1}\int^p_{p_{\rm min}}sF_n(s)ds,\label{eq:B10c}
 \end{equation}
 while combining (\ref{eq:B9}) and (\ref{eq:B10}) 
 \begin{equation}
 \sum_{l=0}^\infty\sum_{n=1}^{l+1}(nZ_ny_{l+1-n,p}-(l+1-n)Z_{n,p}y_{l+1-n})(b-b_{\rm min})^l=\sum_{l=0}^\infty F_l(b-b_{\rm min})^l,
 \end{equation}
 which yields the following upper triangular set of algebraic equations for the coefficients $Z_n$ of the series for $Z$
 \begin{equation}
 \sum_{n=1}^{l+1}(nZ_ny_{l+1-n,p}-(l+1-n)Z_{n,p}y_{l+1-n})=F_l.
 \end{equation}
The equation for the first term  
\begin{equation}
Z_1y_{0,p}=F_0
\end{equation}
 can be readily solved using (\ref{eq:B10c})
 \begin{equation}
 Z_1=\frac{f}{p},
 \end{equation}
 while the other terms satisfy the following recurrence relation
 \begin{equation}
 Z_{l+1}=[(l+1)y_{0,p}]^{-1}\left[lZ_{1,p}y_l-\sum_{n=2}^l(nZ_ny_{l+1-n,p}-(l+1-n)Z_{n,p}y_{l+1-n}) \right],\,l\geq 1.\label{eq:B9.9}
 \end{equation}
 Note how $Z_{l+1}$ depends on the $y_n$'s with $n\geq 1$.  
 
Having determined the map from natural to physical coordinates, we now calculate the kinetic energy of the ground state, by solving  
 the thermal wind equation (\ref{eq:B8.2}), which expressed in $(b,p)$ coordinates reads
 \begin{equation}
 (\sqrt{2Ek})_{,p}y_{,b}-(\sqrt{2Ek})_{,b}y_{,p}=f^{-1}Z_{,p}.\label{eq:B10.1}
 \end{equation}
 Once again, we look for a power series solution
 \begin{equation}
 \sqrt{2Ek}=\sum_{l=0}^{\infty}v_l(p)(b-b_{\rm min})^l,
 \end{equation}
and, since we have assumed that $b=b_{\rm min}$ is the level of no-motion, $v_0=0$. 
Substituting into (\ref{eq:B10.1}) and rearranging we again obtain a triangular algebraic set of equations for the $v_l$'s (no surprise here, as we are dealing with the same nonlinear operator)
\begin{subequations}
\begin{eqnarray}
v_1=0,\\
v_{l+1}=\frac{\left[ -Z_{l,p}+f\sum_{n=1}^l (nv_{l+1-n,p}y_n-(l+1-n)v_{l+1-n}y_{n,p})\right]}{[f(l+1)y_{0,p}]},l\geq 1.
\end{eqnarray}
 \end{subequations}
As before, the $v_{l}$'s with $l\geq 3$ depend on the $y_n$'s with $n\geq 1$. 
In particular,  
 \begin{equation}
 fy_{0,p}v_2=\frac{f}{2p^2}.
 \end{equation}
 An  special case is when  $F_n=0,\,n>0$, i.e. the VDF is linear in the buoyancy. In this case, the solution has the very simple form
 \begin{equation}
 y_*=f^{-1}\int^p_{p_{\rm min}}sF_0(s)ds,\,Z_*=f\frac{b-b_{\rm min}}{p},\,\sqrt{2Ek_*}=\frac{f(b-b_{\rm min})^2}{2p^2}[pF_0(p)]^{-1}.\label{eq:B10.2}
 \end{equation}
 
 With the complete solution in hand, it is now time to revisit the small-${\rm Ro}$ assumption made at the beginning. 
 The goal is to characterize the MDFs that lead to small-{\rm Ro} ground states.  
 It is straightforward to calculate the relative vorticity of the ground state 
 \begin{equation}
 \Omega_*=-(2Ek)_{,b}dtdb+(2Ek)_{,p}dpdt.
 \end{equation}
 Let us start from the special case  that leads to (\ref{eq:B10.2}). The most stringent condition is   
 \begin{equation}
 {\rm Ro}=O\left(\frac{(\sqrt{2Ek_*})_{,Z}}{f}\right)=\frac{L_z}{f^2pF_0(p)}\ll 1
 \end{equation}
 We can estimate 
 \begin{equation}
O(pF_0)=\overline{p}\frac{L_yL_z}{\Delta p\Delta b}=\overline{p}\frac{L_y}{\Delta pN^2},
 \end{equation}
 where $N^2=\Delta b/L_z$. Here $\Delta p$ is a measure of the spread of potential vorticity, e.g. the rms of the fluctuations,  $\Delta b$ is a measure of the spread of buoyancy and $\overline{p}$ the mean potential vorticity (which, when needed, can be estimated as $fN^2$).   
 Thus, the ground state satisfies the low-${\rm Ro}$ condition if the following condition on the spread of PV 
 \begin{equation}
 \frac{\Delta p}{\overline{p}}\ll \frac{L_yL_z}{L_R^2},\label{eq:B10.4}
 \end{equation}
 where $L_R\equiv L_zN/f$ is the internal Rossby radius of deformation, is satisfied. 
 
 Let us now consider the more general case of a MDF which is not linear in the buoyancy, but which still satisfies (\ref{eq:B10.4}). We have
 \begin{equation}
 O(pF_n)={\rm Ro}^{-1}\frac{L_R^2}{\Delta b}(\Delta b)^{-n},
 \end{equation}
 whence
 \begin{subequations}
 \begin{eqnarray}
 O(fy_n)=O(\Delta p\, pF_n)=fL_y(\Delta b)^{-n},\label{eq:B11a}\\
 O(fy_{n,p})=O(pF_n)={\rm Ro}^{-1}\frac{L_R^2}{\Delta b}(\Delta b)^{-n}.
 \end{eqnarray}
 \end{subequations}
Using (\ref{eq:B9.9}), we have 
\begin{equation}
O(Z_2(b-b_{\rm min})^2)={\rm Ro}\frac{L_yL_z}{L_R^2}L_z.
\end{equation}
Given the nature of the recurrence relation (\ref{eq:B9.9}), it  follows that all the other terms in the series 
for $Z$ are $O({\rm Ro})$,  and thus 
\begin{equation}
Z_*=\frac{f}{p}(b-b_{\rm min})+O({\rm Ro}).\label{eq:B11}
\end{equation}
Along essentially the same lines we have that  $\sqrt{2Ek_*}=O({\rm Ro})$, and thus to $O({\rm Ro})$ the pressure in the ground state is hydrostatic. 
Hence
\begin{equation}
\overline{\Psi}=-\frac{f}{2}\frac{(b-b_{\rm min})^2}{p}+O({\rm Ro}),\label{eq:B12}
\end{equation}
which shows that the Casimir of geostrophic leaves, defined as leaves characterized by a geostrophic ground state (i.e., a low-${\rm Ro}$ state)  does not depend, to leading order, on the details of the MDF that defines the leaf to which the ground state belongs. 
A field on a geostrophic leaf does not need to be in geostrophic equilibrium, but the ageostrophic component must be such that the overall spread in potential vorticity is small when looked over the entire field.    
 As long as the diabatic dynamics does not cause a significant spread in potential vorticity, the flow remains on a geostrophic leaf and the gauge-fixed Hamiltonian is to leading order independent on the MDF. Thus, the local diagnostic energy has, to $O({\rm Ro})$ a universal character. 
 \subsection{The non-holonomic dynamics of Casimirs in Boussinesq flows}

Once the gauge-fixed Casimir is known, non-holonomic effects on the Available Energy can be analyzed by considering how the Casimir evolves along Lagrangian trajectories. In apedic manifolds characterized by an even distribution of potential vorticity, {our} theory recovers the standard SBB local APE formulation (here, we reinstate the qualified Potential, since the ground state has no kinetic energy). In particular, under Boussinesq approximation, the sink of APE is quantified by $-\kappa \overline{\psi}_{,bb}|\nabla b|^2$, where $\kappa$ is the diffusivity of the stratifying agent. Archetypal flows for small-scale mixing (e.g., shear layers, Couette flows,\ldots) fall into this category. On geostrophic manifolds, the evolution of the Casimir depends on the details of the non-holonomic brake. Assuming a standard diffusion term in the buoyancy equation, the sink term due to mixing of the stratifying agent $-\kappa \overline{\psi}_{,bb}|\nabla b|^2$ is equal to $\kappa f/p|\nabla b|^2+O({\rm Ro})$. However, turbulent momentum fluxes will change the potential vorticity along trajectories and thus modify the Available Energy in either direction, just as we saw for the shallow-water case. The specific way depends in general on how the turbulent fluxes are modeled, and details are left as future work. 
\section{Relation to other definitions of local and global Available Energy}
 Examples of global \citep[e.g.,][]{WintersLRD95}  as well as local Available Energy definitions \citep[see, e.g.,][]{KucharskiT00,CodobanS03,Andrews06,ScottiW14} applied to study the effects of mixing on the energetics of three-dimensional flows have been previously considered in the literature. 

At the level of a global definition of Available Energy, the novelty of the approach followed in sec.~\ref{sec:theory} lies in setting a framework that allows, at least in principle, to calculate the ground state from the MDF that describes the state of a generic state, though it can also be used to calculate the MDF associated to a prescribed ground state. While the framework is amenable to analytic treatment in some special cases, it suggests a way to calculate numerically the ground state via the application of suitably defined holonomic brakes. Once the ground state is known, the gauge-fixed Casimir can be calculated, and thus the local diagnostic energy is obtained. 
Out of the cases that can be treated analytically, we show that for an important class of flows, the  isochoric restratification proposed by Winters et al. coincides with the  ground state obtained with our framework.

The major departure from previous local formulations of APE 
\citep[e.g.,][]{ScottiW14}, aside from the way the Casimir is calculated, resides in not subtracting the contribution of the ground state at the local level. If we were to follow Scotti and White's "recipe", we would define the local diagnostic energy as ${\cal E}^{\rm SW}(q)={\cal H}_*(q)-{\cal H}_*(q_*)$, with ${\cal H}_*$ given by the gauge-fixing condition (\ref{eq:E4}).

Subtracting the gauge-fixed Hamiltonian density of the ground state  in the local definition  has an intuitive appeal, because the global Available Energy would coincide with ${\cal E}^{\rm SW}[q]$, whereas in the approach followed here the integral of the local diagnostic energy does not have an immediate physical interpretation. However, the quantity of interested is the rate of change of the Available Energy under diabatic conditions, which coincides with the rate of change of  ${\cal E}[q]$. Thus, the local diagnostic energy allows to explore at the local level the processes that are responsible for energy degradation, and at the local level there are reasons to question the inclusion of the ground state energy in the local definition. 

From a functional point of view, the inclusion of ${\cal H}_*(q_*)$ amounts to no more than a "constant" term, in the sense that its variation w.r.t. $q$ is zero, and it is not necessary to impose either the gauge condition or to ensure   the convexity property.

There is also a subtler reason why it is not advisable to include $q_*$ in the definition at the local level.  Consider the local evolution of ${\cal H}_*(q_*)$ under the flow $\bm v$ associated to the point $q$ in phase space. For simplicity, we consider the stratified  Boussinesq case, where the volume form is an integral invariant. Then ${\cal H}_*(q_*)={\cal H}_*(\lambda_*,b_*,Z)$ (we do not need to worry about the explicit time dependence of ${\cal H}_*$) and 
\begin{equation}
\begin{split}
 \frac{\partial({\cal H}_*\mathfrak{V})}{\partial t}+{\cal L}_{\bm v}({\cal H}_*\mathfrak{V})=
 \frac{\delta ({\cal H}_*\mathfrak{V})}{\delta \lambda_*}\left(\frac{\partial\lambda_*}{\partial t}+{\cal L}_{\bm v}(\lambda_*)\right) \\
+\frac{\delta ({\cal H}_*\mathfrak{V})}{\delta b_*}\left(\frac{\partial{b}_*}{\partial t}+{\cal L}_{\bm v}(b_*)\right)
+\frac{\delta ({\cal H}_*\mathfrak{V})}{\delta Z}{\cal L}_{\bm v}(Z)
\cong \frac{\delta ({\cal H}_*\mathfrak{V})}{\delta Z}{\cal L}_{\bm v}(Z),
 \end{split}\end{equation}
where, as before, $\cong$ means equal up to an exact form. For an apedic manifold, 
\begin{equation}
\frac{\delta ({\cal H}_*\mathfrak{V})}{\delta Z}{\cal L}_{\bm v}(Z)=-b_*(Z)d(Z\Phi)=-d(P_*(Z)\Phi)\cong 0,\,\Phi=i_{\bm v}\mathfrak{V},\,P_*=\int^Z b_*(s)ds,
\end{equation}
so that in the adiabatic limit, ${\cal E}^{\rm SW}$ is, up to boundary terms, an integral invariant.  

However, under more general ground states (e.g., geostrophic), ${\cal E}^{\rm SW}$ is not an integral invariant, whereas ${\cal E}$ is. Therefore, referencing the local diagnostic energy to the local background state does not contribute to the desirable properties of the local diagnostic energy, while, aside from special circumstances, adding terms to the local evolution equation of dubious physical interpretation.

\section{Conclusion}


In this paper, we have developed a framework to diagnose  diabatic effects on systems that in the adiabatic limit are described by a degenerate Hamiltonian structure. The degeneracy has two important consequences. The first is that the phase space foliates into a collection of leaves, each labeled by a particular volumetric distribution of the Lagrange invariants associated to the Casimirs. Trajectories in phase space are contained within a leaf. The Available Energy on a leaf is thus the energy referenced to the minimum energy attainable {\em on the leaf}. In other words, it is the energy that can be extracted from the system without disturbing the dynamics of  the Lagrangian invariants. 

The second consequence is that the Hamiltonian that describes the flow in the adiabatic limit possesses a local (in phase space) gauge symmetry. Thus, there exists a class of dynamically equivalent Hamiltonians. 

We show that a specific gauge-fixing condition can be imposed to select a specific Hamiltonian that can be used to define a local diagnostic energy with the following property: the temporal rate of change of the Available Energy  under diabatic conditions is given by the diabatic evolution of the gauge-fixed Hamiltonian.

The present study provides expressions of the gauge-fixed Hamiltonians for a number of geophysically relevant flows, together with general properties of the reference state. 
For non-rotating flows in the Boussinesq limit, we have shown that the local diagnostic energy based on the gauge-fixed Hamiltonian recovers the Available Potential Energy of \citet{HollidayM81}, which form the basis of \citet{WintersLRD95} and \citet{ScottiW14} application of APE to mixing, provided the distribution of potential vorticity in the flow is even around the origin. Most archetypal flows used to study small-scale mixing start as two-dimensional flows contained in a vertical plane. For such flows, under the assumption that the diabatic dynamics does not favor negative over positive potential vorticity,  the standard  approach remains correct.

Our definition of local diagnostic energy extends naturally to rotating flows: under the effect of rotation,  the conservation of potential vorticity imposes a strong constraint, which is not accounted for in the Winter et al.'s definition of APE. In particular, when the flow is obtained from a general-form perturbation  which preserves both the potential vorticity and the buoyancy distribution of a low-Rossby number state in near-geostrophic equilibrium,
to lowest order in the Rossby number, we show that the expression for the local diagnostic energy has a universal character.

Once the functional form of the local diagnostic energy is known, it can be used to study the effect of diabatic processes. The effect of these processes, which in general  break the conservation of the Lagrangian invariants on the change of the Available Energy, can be studied  following the evolution of the local diagnostic energy along Lagrangian trajectories.
Using a simple model for diabatic processes in  shallow-water flows, we have shown that such processes can locally increase the Available Energy. 
This is not too surprising, since diabatic processes erode the potential vorticity constraint, keeping energy locked up in the ground state field. 

The results were derived under the assumption that the ground state is unique. Of course, we can envision situations where the naive Hamiltonian has multiple minima on a given leaf. In this case, it cannot be ruled out that the application of different holonomic brakes may lead the different ground states, especially if the system is close to the separatrix between the basin of attractions. On the other hand, if the system is not too far from a minimum, then it is not unreasonable to assume that most holonomic brakes will lead to the same extremal point, even though there may exists other ground states with overall less energy.    

We considered applications of the general formalism developed in section ~\ref{sec:theory} to flows in which thermodynamic effects are neglected. 
However, the formalism is quite general, and can be applied to any flow regime described by a Hamiltonian on a phase space with a degenerate Poisson algebra. In particular, it could be applied to flows where the internal energy plays a more active role, such as low Mach number applications \citep{emanuel:94,holloway:09,Pauluis:15}.

In conclusion, the formalism  presented in this paper extends Margules' original idea to highlight the role that multiple constraints play in the available energetics framework.

Since the Mass Distribution Function, or, for incompressible flows, the Volume Distribution Function plays a central role in defining the gauge-fixing condition, our approach applies to flows contained on a finite volume. 
 However, the possibility that other ways to enforce the constraints, not dependent on the system having a finite volume, exist should not be discarded. 
Since  the applicability of any local diagnostic tool is constrained by the data available for the analysis, it can rarely be considered independent from the flow data itself 
\vskip6pt

\enlargethispage{20pt}

\ethics{The research described in this paper did not involve any living or deceased organism.}

\aucontribute{AS conceived the study and drafted the manuscript. AS and PYP worked together on the calculations, PYP heavily edited the manuscript and contributed some of the figures. }

\competing{We have no competing interests.}

\funding{The authors acknowledge the support by the National Science Foundation Grant N$^o$ OCE-1155558.}

\ack{We would like to thank Dr. K. Lamb, Dr. A. Hoggs and Dr. R. Tailleux for numerous engaging discussions on the subject. 
AS would like to thank Dr. E. Santilli for opening his eyes to the beauty of exterior calculus. }

\disclaimer{Insert disclaimer text here.}

\bibliographystyle{rspublicnat}
\bibliography{local.bib,internal_waves.bib}

\end{document}